\RequirePackage{fix-cm}
\documentclass[smallextended]{svjour3}       % onecolumn (second format)

\smartqed  % flush right qed marks, e.g. at end of proof
\usepackage{graphicx}
\usepackage{amsmath}

\newcommand{\ket}[1]{|#1\rangle}
\newcommand{\up}{\!\uparrow}
\newcommand{\down}{\!\downarrow}
\journalname{Foundations of Physics}

\begin{document}

\title{Perspectival Quantum Realism}

\titlerunning{Perspectival Quantum Realism} 

\author{Dennis Dieks}

\authorrunning{Dennis Dieks}

\institute{D. Dieks \at
              History and Philosophy of Science, Utrecht University, Princetonplein 5, 3584CC Utrecht, The Netherlands\\
              \email{d.dieks@uu.nl} } 

\date{Received: date / Accepted: date}

\maketitle

\begin{abstract}
The theories of pre-quantum physics are standardly seen as  representing physical systems and their properties. Quantum mechanics in its standard form is a more problematic case: here, interpretational problems have led to doubts about the tenability of realist views. Thus, QBists and Quantum Pragmatists maintain that quantum mechanics should not be thought of as representing physical systems, but rather as an agent-centered tool for updating beliefs about such systems. It is part and parcel of such views that different agents may have different beliefs and may assign different quantum states. What results is a collection of agent-centered perspectives rather than a unique representation of the physical world.

In this paper we argue that the problems identified by QBism and Quantum Pragmatism do not necessitate abandoning the ideal of representing  the physical world. We can avail ourselves of the same puzzle-solving strategies as employed by QBists and pragmatists by adopting a \emph{perspectival quantum realism}. According to this perspectivalism (close to the relational interpretation of quantum mechanics) objects may possess different, but equally objective properties with respect to different physically defined perspectives. We discuss two options for such a perspectivalism, a local and a nonlocal one, and apply them to Wigner's friend and EPR scenarios. Finally, we connect quantum perspectivalism to the recently proposed philosophical position of \emph{fragmentalism}. 

\end{abstract}

\section{Introduction}\label{intro}

Pre-quantum physical theories are standardly interpreted as providing descriptions of physical systems and their properties; classical instrumentalists and skeptics typically admit the existence of such realist stories before rejecting them or suspending belief. But in quantum physics interpretative problems have created doubts about the general tenability of a realist outlook. In particular, QBists and Quantum Pragmatists maintain that standard quantum mechanics is not representational but rather has the status of a calculus enabling us to update our beliefs about future experiences. These agent-centered approaches succeed in dissolving a number of puzzles, in particular relating to Wigner's friend and EPR scenarios. 

In this paper we address the question of whether the drastic step of abandoning the ideal of representation is necessary; whether it might be possible to retain the puzzle-solving successes of non-representational strategies within a realist context. 

But first, what should count as ``standard quantum mechanics''? Textbooks typically follow the tradition of presenting quantum theory as comprising two principles of dynamical evolution: unitary evolution when no measurements are taking place and ``collapses of the wave function'' in the case of measurements. 

It has long been recognized that this dynamical non-uniqueness leads to serious conceptual problems. In particular, it is responsible for ambiguity concerning the domains of validity of the two evolution principles.\footnote{It might be noted here that even some of the founding fathers often associated with the introduction and defense of the notion of collapse, in particular von Neumann and Bohr, were not unequivocal about its status as a dynamical principle, see \cite{dieksbohr,diekspersp}.} This ambiguity is at the basis of the quantum measurement problem and other interpretative issues, \emph{e.g.}\ those relating to locality and the paradox of Wigner's friend. The absence of a clear dividing line between the domains of applicability of the two types of evolution may even be considered a threat to the formal consistency of quantum mechanics. 

We might therefore expect confusion in the practice of quantum physics. But there isn't: the question of when exactly collapses take place is insignificant in actual practice. There exists broad agreement in the quantum community that what counts for applications is the prediction of probabilities (including those for repeated measurements, and conditional probabilities), and that the unitary formalism---without collapses---is in principle sufficient for this purpose (\emph{cf}.\ \cite{healeyobj}). 

That unitary evolution and collapses do not possess the same  significance receives further support from a growing list of experiments explicitly verifying the existence of superposed states of mesoscopic and near-macroscopic systems. These experiments suggest that linear evolution is never suspended, not even in interactions with macroscopic objects---although it is true that due to decoherence macroscopic superpositions are very difficult to detect. 

This self-sufficiency of the unitary formalism, together with its consistency (in contradistinction to the questionable consistency of ``unitary plus collapse quantum mechanics'') motivate considering unitary evolution as the sole dynamic principle of quantum theory. Accordingly, our aim here is to focus on the interpretation of unitary quantum mechanics, in which collapses do not represent independent dynamical physical processes (but may have another interpretational role to play). 

In unitary quantum mechanics all evolution is linear, so that the final state after a measurement interaction will be a superposition to which all possible measurement outcomes contribute; the actually observed post-measurement situation is not singled out. This presence of superpositions where single terms may be expected is one reason for doubting that the unitary formalism can be descriptive of physical reality.

This doubt gains strength when ``nested'' experiments of the Wigner's friend (WF) type are considered. In WF setups an experimenter in a sealed laboratory (the friend) successfully performs a quantum measurement and finds a definite result. If this outcome is the only piece of information that the friend possesses, it is natural for her to use the eigenstate corresponding to the obtained outcome for predictive purposes. However, an outside observer (Wigner), who knows that a measurement is taking place inside but has no access to its outcome, will use unitary evolution for his calculations and will therefore base his predictions on a superposed state in which his friend's actual result has no privileged status. Cases like this may suggest that quantum mechanics is a ``single user theory'',  yielding different results for different agents rather than leading to a coherent representation of the physical world. 

If this line of thought is consistently pursued, the ideal of a universally valid quantum description of the physical universe becomes a chimera. Indeed, according to QBists it should not be considered the ambition of quantum theory, or for that matter physics in general, to develop a unique and global picture of the physical world. Rather, QBists maintain, the quantum states that are assigned to physical systems, the Hamiltonians that are judged to apply, and the probabilities that are calculated represent beliefs of different agents (like Wigner and his friend) using the theory. Thus, the quantities figuring in the quantum formalism do not describe physical systems but represent epistemic states of users of the theory.        

\section{Quantum mechanics as an agent's personal probability theory}\label{qbism}

QBists argue that  quantum mechanics provides agents with a users manual for how to form rational expectations about the consequences of their interventions in the external world. Inherent in this doctrine is the interpretation of quantum probabilities as \emph{subjective}. That is, QBist probabilities do not reflect relative frequencies, objective chances, or some other notion of physical probability; they rather serve to quantify personal, subjective degrees of belief.  

The subjective nature of QBist probabilities is illustrated by the meaning given to probability-$1$ statements. If a QBist agent predicts an experimental result with probability $1$, this does not imply anything about the physical status of that  future result; in particular, it does not entail that the result will necessarily be realized, or that the result in question is already there in the external world, waiting to be revealed. The only implication is that the agent is completely convinced that the outcome in question will be found. This is a fact about her or his expectations, not about the physical world. Another agent, with different information and/or different beliefs, may consistently have different expectations and may therefore assign different states leading to other probabilities. QBism thus leads to a \emph{fragmented} picture, consisting of different subjective perspectives.

\subsection{Wigner and his friend}\label{qbism2}

The WF scenario described in the Introduction illustrates how these agent-centered ideas work.  Wigner's friend carries out a measurement, takes cognizance of its result and, being certain about this result, assigns a probability $1$ to it. This seems analogous to what happens in classical situations of lack of certainty, when the uncertainty is lifted; for example when we observe that a particular ball is drawn from an urn.   In quantum mechanics, however, a series of no-go results stands in the way of interpretations in terms of becoming aware of a preexisting reality. Therefore, QBism asserts that the assignment of probability $1$ after a measurement should not be seen as reflecting certainty about a previously existing state of affairs, but rather as certainty about a novel situation created by the agent's intervention. 

Wigner and his friend are different agents who, from the QBist perspective, constitute different ``centers of subjectivity'' from which the world is contemplated. They assign different quantum states to their individual external worlds and make different predictions.  

Wigner, who receives no information from inside the lab but knows that his friend will perform a measurement, will use unitary quantum mechanics to update the initial state he has assigned to the lab and its contents. He will thus arrive at a superposed state in which terms corresponding to different internal measurement results occur. This state will form the basis for Wigner's expectations concerning later measurements that he could undertake. \footnote{Probabilities of consecutive measurements outcomes, and conditional probabilities, can also be predicted from the unitary formalism. Thus, instead of arguing that a state $|\Psi\rangle = \Sigma c_i |\phi_i\rangle$ will collapse into some $|\phi_k \rangle$ in measurement $1$, and then asking for the probability that in subsequent measurement $2$ the result corresponding to $|\phi_j \rangle$ will be found in this state $|\phi_k \rangle$, one may find joint and conditional probabilities by calculating the expectation value of $ | \phi_k\rangle \langle \phi_k | \otimes   | \phi_j\rangle \langle \phi_j | $ in the entangled state of system and measuring devices that results from unitary evolution.} 

Wigner's friend obtains information that is  not available to Wigner: she finds a definite outcome of her measurement, she updates her personal probability for that outcome to the value $1$, and assigns a corresponding quantum state to the system and her lab.  

Wigner's friend will consequently work with a quantum state that differs from the state assigned by Wigner. Wigner assigns a superposition to the total lab system, and therefore a mixed state to the system measured by his friend; in this mixed state all possible measurement outcomes are represented. By contrast, Wigner's friend assigns an eigenstate of the measured observable, corresponding to the single measurement result she has actually registered.

That the theoretical perspectives of Wigner and his friend are different is not at all surprising from the QBist point of view. Indeed, it is essential for QBism that the different assigned states are not rival descriptions of one and the same physical system, competing for  objective truth. Rather, they represent the different personal opinions of two different agents. Given that these agents possess divergent pieces of information and find themselves in different predicaments, with different possibilities of intervening in the external world, it is only natural that their expectations differ.

\subsection{Locality}\label{qbism3}

According to QBism quantum mechanics pertains solely to what goes on in an agent's mind. This has the immediate consequence that quantum mechanics is \emph{local}. Indeed, all quantum predictions are about what happens at the position of the agent, namely the agent's local perceptions and expectations. QBist quantum mechanics never leaves this local domain; in particular, it never makes statements about what objectively happens at a distance, in the external world. Quantum mechanics in the QBist view therefore never commits itself to statements about non-local correlations between events with space-like separation. 

The QBist analysis of the Einstein-Podolsky-Rosen thought experiment illustrates this locality. When Alice performs a measurement and obtains a definite result, she will adjust her credences. She will assign probability $1$ to one particular result that she expects to find at the other side of the experiment, if and when she travels there and measures Bob's particle. This change in Alice's expectation does not change anything physical at Bob's location; in particular, it does not entail that the predicted property exists and already existed previously at Bob's side (as originally argued by EPR). The only implication of Alice's probability updating, according to the QBist, is that Alice has become convinced that she will verify the correlated result when she measures Bob's particle. This is a local and instantaneous change in Alice herself. If Alice later gets hold of Bob's particle and actually measures it, and really finds her predicted result, it does not follow that this result was there all along, waiting to be found. As discussed a moment ago, measurements must be considered to \emph{create} outcomes, not to reveal them. So, the whole story is about events occurring at Alice's position;  nothing non-local enters the account. An analogous QBist story can be told from Bob's perspective. 

The success of the QBist strategy for salvaging locality depends on the assumption that quantum mechanics is about different agents reasoning within different personal perspectives. However, as we will argue in section \ref{pqr}, key elements of this QBist strategy can be co-opted if a perspectival \emph{realism} is adopted. Before venturing in this realist direction it is helpful to have a critical look at  the central role of subjectivity in QBism.      

\section{From subjective to objective expectations}\label{obj}

According to QBism quantum mechanics affords us a probability calculus by means of which we can update our subjective beliefs and expectations. This emphasis on the private nature of what quantum mechanics is about may appear to smack of solipsism. 

But QBists emphatically reject solipsism: they argue that the very purpose of their approach is to arrive at expectations concerning \emph{the external world}. Moreover, QBists acknowledge \cite{fuchsWFloc,fuchsreal,healeysep} that the proven success of the quantum formalism in dealing with the external world must reveal something objective about what this external world is like.\footnote{As one such objective feature QBists cite the ``creative nature'' of measurements: measurements do not reveal preexisting properties but create novel features.}

Indeed, the undisputed helpfulness of quantum mechanics for finding our way in the physical world makes it hard to deny  that the rules of the quantum calculus in some way latch on to the structure of the world. This admission creates a tension, however, with the idea that quantum mechanics is purely about subjective states of belief, quantified by equally subjective probabilities. In applications of the theory certain assignments of states and probabilities certainly prove to be more successful than others, as demonstrated every day in laboratory practice.

As Healey points out \cite{healeyobj}, there exists widely shared agreement in the quantum community about how to assign quantum states in specific situations. It is true that differences of opinion can nevertheless arise, but there are well-established norms in physical practice for how to discuss and eventually resolve such disagreements. These norms are not arbitrary but relate to the scientific goals of accurate prediction and explanation. Eventual consensus about which predictions are valid is essential for designing and controlling experiments and for technical applications---and for the well-functioning of science in general.

For this reason Healey has proposed 
\cite{healeyprag,healeysep,healeyrelativism,healeyobj} what might be called an ``objectified'' version of QBism, Quantum Pragmatism (QP), in which the assignment of quantum states is not a purely subjective decision and in which Born probabilities are taken to be objective. In this proposal quantum mechanics is still non-representational and agent-centered: it does not provide users with a description of the physical world, but it gives \emph{normative advice} about how to deal with that world. Accordingly, Quantum Pragmatists distinguish between ``quantum claims'', about quantum states and probabilities, and ``non-quantum magnitude claims''. The latter are about physical systems and the values of physical quantities (\emph{e.g}., position, momentum and spin), and are descriptive in the ordinary sense. By contrast, quantum claims are \emph{prescriptive}: they supply \emph{norms} for the use of descriptive non-quantum claims and the degrees of belief a user should attach to them. 

According to Healey, \emph{decoherence} plays an important role in \emph{legitimizing} making descriptive statements. When decoherence has occurred, the Born rule assigns consistent Kolmogorov probabilities to the values of all physical quantities in whose eigenbases  the system's density operator (obtained by partial tracing) has become diagonal. Healey therefore proposes that in such cases decoherence licenses the use of descriptive statements about these quantities. For example, if the density operator has become diagonal in the position basis, we are justified in asserting statements of the form ``the position has value $x$'', and the Born rule prescribes the degree of belief $p(x)$ we should have in this proposition.   

This pragmatist move is a significant step away from QBist subjectivity. The picture that now arises is still fragmented: different users of quantum mechanics may assign different quantum states to their respective external worlds. But according to quantum pragmatism there are good physical reasons for these assignments, and because of these reasons physicists are committed to their predictions. These predictions should be   verifiable by standard physical techniques, and in this sense aim at objective correctness. 

The Wigner's friend scenario can again be used as an illustration.  Wigner is licensed to make objective descriptive predictions about the outcomes of measurements of decohered quantities pertaining to the sealed lab. If he follows the quantum rules correctly (and if quantum mechanics itself is correct), these predictions will be born out by experiment. Wigner's friend makes equally justified predictions about what will happen inside her lab, in terms of quantities that have decohered for her---she has no access to the total coherent quantum state. The states assigned by Wigner and his friend are thus different but both motivated on physical grounds and thus objective within their own respective perspectives.

Like his QBist colleague, the quantum pragmatist warns us that we should not think that quantum states represent and describe physical systems the way they are \emph{per se}. Rather, the states establish agent-centered norms for which statements about the world are appropriate and for fixing the faith agents should put in these statements. 

Summing up, both QBism and Quantum Pragmatism make agents\footnote{In \cite{healeyrelativism} Healey explains that an agent need not necessarily be a conscious user of the theory; an inanimate physical system interacting with the world in a situation where decoherence conditions are fulfilled can also be an agent. This makes the difference with realist interpretations even smaller than it already was---see further in the main text.\label{agent}} the focal points of their discussions of what quantum mechanics is about. In QBism this picture is based on the conception that all quantum predictions represent agents' subjective states of belief. Quantum Pragmatists object that this position is irreconcilable with the general physical principle that what matters for prediction and explanation is not what agents subjectively believe about physical processes, but rather what they \emph{should} think about them; the dominant role of subjectivity in QBism is at odds with the practice of physics. QP accordingly takes states and probabilities to be objective, in the sense of latching on to the physical reality that reveals itself to an agent. QP thus maintains that state attributions are centered on agents, and that these states, and the quantum formalism \emph{in toto}, do not \emph{describe} physical systems but provide rules for arguing about them and making predictions. 
 
This emphasis on agents and the non-representational character of the theory makes it possible for QP to adopt the same strategies as QBism in the contexts of Wigner's Friend and the locality question. In the former case the pragmatist argues that Wigner and his friend, in their different physical predicaments, arrive at different objective predictions. If they assign different states to the same system this is no contradiction because these states are not meant to represent the system. Concerning the locality issue the pragmatist argues, as the QBist, that all quantum predictions are about what happens locally to an agent, so that the problem of non-local correlations between systems at a distance does not arise. 
 
This summary shows that there is a gap between QP's agent-centered normative reasoning about the world and a representational realist account; but it also suggests that the gap is bridgeable. For example, in Healey's pragmatist view it is essential that Born probabilities are objective and normative. It is true that these probabilities are interpreted as guiding an agent's degrees of belief, but these beliefs are meant to prove objectively correct after comparison with the outcomes of (repeated) experiments, with the help of standard statistical methods. Further, although quantum states are stated to be non-representational, their form (decohered or not in some basis) is considered decisive for the question of which descriptive statements can legitimately be made. It is a relatively minor step to go from this to the attribution of a \emph{descriptive} role to states: if states have been decohered in a certain basis, they \emph{represent} a definite-valued property of the system (namely the property corresponding to the  observable whose eigenbasis is singled out). 

In the next section we will explore this option of  a representational, realist interpretation of unitary quantum mechanics, while retaining perspectivalism and the related puzzle-solving strategies of QBism and QP.

\section{Perspectival quantum realism}\label{pqr}

Realism should not focus on agents and their subjective points of view; if perspectivalism is to be part of a realist scheme, the perspectives in question should be defined with respect to physical systems. This implies a transition from a collection of agent-centered views to an account in terms of different physical perspectives, centered on different physical systems. An agent in the sense of someone  who intervenes in the world then becomes a special case of a physical system interacting with its environment.\footnote{As mentioned in footnote \ref{agent}, Healey's pragmatism already incorporates this more general notion of an ``agent'' as a physical system fulfilling certain conditions.} 

Unitary quantum mechanics has the essential characteristic that it preserves superpositions and generally leads to the formation of entangled states during interactions. This is true even in the case of interactions  with macroscopic measuring devices, which leads to the measurement problem. QBism and Quantum Pragmatism evade this problem by denying superpositions a descriptive role; according to these interpretations superposed quantum states only yield probabilities, to be used by agents. That the experiences of these agents are unique, consisting in exactly one perceived state of affairs, is treated as an unproblematic basic fact. 

But this strategy is not limited to QBists and pragmatists: realists can adopt it as well, with the note that they will not, of course, assign a privileged status to human perceptions and experiences. The crux of this realist strategy is to assume that quantum states provide an objective  \emph{probabilistic} description of the world. On this assumption a post-measurement superposition of object and device can be regarded as representing a range of possibilities, one of which will actually obtain. A probability should be specified for each individual possibility to be realized. 

In the context of a realist interpretation it is natural to think of objective and physical probabilities. This notion of objective probability may be fleshed out in a variety of ways: for example, in terms of a Humean best theory explanation or Lewisian objective chances, or perhaps with an appeal to relative frequencies of some sort. For our purposes here the details of how objective probabilities should be analyzed are not important. Of course, whatever the analysis, there should be a relation to the frequencies of results in long series of experiments, and as soon as agents in the ordinary sense appear on the scene they should do wise to use objective probabilities as a guide for their expectations. The Qbist and QP recommendations for how to use quantum mechanics should have a natural place within the realist scheme.  

Several proposals for such probabilistic realist interpretations of unitary quantum mechanics exist in the literature. One example that springs to mind derives from the many-worlds interpretation, according to which each individual term in a superposition represents a definite state of affairs, characterized by well-defined values of some set of observables. It is true that the many-worlds interpretation considers all terms in a superposition as representing actualities. However, this ontology of many coexisting worlds does not do practical work, since our epistemic access is restricted to one single world. When we confine our attention to the physically accessible part of the many-worlds ontology, namely the world we live in, what remains is a single-world interpretation of unitary quantum mechanics making probabilistic predictions. 

Seen in this way, the many-worlds interpretation affords us an example of a \emph{modal} interpretation of quantum mechanics. Modal interpretations start from the idea that quantum mechanics makes probabilistic statements about a single world. The assumption is that in an entangled state of the von Neumann measurement type only one ``branch'' corresponds to what is actual; the other branches correspond to unrealized possibilities---\emph{modalities}.\footnote{It may well be that Everett himself meant his relative state interpretation \cite{everett}, which inspired the many-worlds ideas, in precisely this way, as a probabilistic single-world theory \cite{barrett-1,barrett0,barrett1,barrett2,barrett3,barrett4,dieksoxfhb}.} 

Some modal interpretations posit the existence of \emph{a priori} preferred observables that are always definite-valued, others assume that the set of definite-valued quantities, and so the set of possibilities, depends on the form of the quantum state and can change over time (see \cite{bub,dieksvermaas,lombardidieks,dieksoxfhb} for overviews). The presently most popular version of the many-worlds interpretation \cite{wallace} uses decoherence, and the formation of stable patterns, as a criterion for definiteness of properties. Translated to the single-world viewpoint, this is a realist counterpart to Healey's criterion saying that decoherence licenses talk about definite quantities. Modal interpretations that employ the spectral decomposition of a system's density matrix in order to define definite-valued quantities \cite{lombardidieks,dieksoxfhb,vermaas} are similar to this decoherence criterion. 

Rovelli's relational interpretation of quantum mechanics \cite{rovelli,rovelli2,laudisa} is another example of an extant single-world probabilistic scheme. This interpretation characterizes the dynamical properties of physical systems as \emph{relational}: a system's properties are defined with respect to other systems.  According to Rovelli physical quantities of two systems become definite-valued, with respect to each other, when an interaction between the two systems correlates the quantities in question (so that there is an ``exchange of information'' between the two systems). Mathematically, the creation of such correlations is expressed by the formation of an entangled state, so that the definition of relational definite-valued quantities in this interpretation follows the pattern of modal interpretations using decoherence or the diagonalization of the density matrices obtained by partial tracing. 
       
More generally, \emph{all} probabilistic single-world interpretations of unitary quantum mechanics that use something like decoherence or diagonalization of density matrices to pick out definite-valued quantities naturally lead to a relational (or perspectival) picture of the world (\emph{cf}.\ \cite{benedieks}). The Wigner's friend scenario shows why this is the case.

\subsection{Wigner's Friend and perspectivalism}\label{qpr2}

Suppose that $A$ (Alice inside her sealed laboratory) measures the spin in the $z$-direction  of a spin-$1/2$ particle that is in the spin state $\frac{1}{\sqrt2}(\ket{\up} + \ket{\down}$. In this measurement the spin states $\ket{\up}$ and $\ket{\down}$ become entangled with $A$'s memory, the register in with the outcomes are recorded, and other parts of the laboratory. Let these macroscopically different total configurations be represented by orthogonal states  $|A_{\up}\rangle$ and $|A_{\down}\rangle$, respectively. Assuming unitary evolution, the state of the composite system ``spin + $A$'s laboratory'' will during the measurement evolve to: 
\begin{equation}
|\Phi\rangle = \frac{1}{\sqrt2} \left(|\up \rangle |A_{\up}\rangle + |\down \rangle|A_{\down}\rangle \right).
\label{eq:wignerstate}
\end{equation}

This state of the laboratory and its contents is a coherent superposition even though a process of decoherence has taken place inside the lab, so that in Alice's internal spin measurements no interference between the two terms in the state will show up. However, outside the lab Wigner, represented by  $W$, can verify that his state assignment~(\ref{eq:wignerstate}) is correct by measuring an observable of which this state is an eigenstate; if we simplify by describing $A$'s laboratory with the help of a two-dimensional Hilbert space spanned by $|\up \rangle |A_{\up}\rangle$ and $|\down\rangle|A_{\down}\rangle$, any observable with the eigenstates  $|\Phi^{\pm}\rangle =\frac{1}{\sqrt2} \left(|\up \rangle|A_{\up}\rangle \pm |\down\rangle |A_{\down}\rangle \right)$ will do the job.  Measurements of such an observable  will yield the outcome associated with  eigenvector $|\Phi^+\rangle$ with probability $1$, which can be checked in a long series of experiments.

Alternatively, $W$ may wish to find out which outcome was found by $A$. To this end, he could measure an observable with the eigenstates $|\up \rangle |A_{\up }\rangle$ and $|\down \rangle|A_{\down }\rangle$. If we denote these states by $|A_{up}\rangle$ and $|A_{down}\rangle$, respectively, and take $\pm 1$ as $W$'s possible measurement outcomes, the two observables between which $B$ has a choice take the form 

\begin{eqnarray}
& \mathcal{B}_1 =  |A_{up}\rangle \langle A_{down}| + |A_{down} \rangle \langle A_{up}|, \\
& \mathcal{B}_2 = |A_{up}\rangle \langle A_{up}| - |A_{down} \rangle \langle A_{down}|,
\end{eqnarray}
respectively. Measuring $\mathcal{B}_1$ provides information about the total entangled state of the laboratory and its contents, which is a superposition of $|A_{up}\rangle$ and $|A_{down}\rangle$; the result of measuring $\mathcal{B}_2$ is usually interpreted as revealing to Wigner whether $A$ has found either ``up''  or ``down''---see, however, our discussion below. 

$\mathcal{B}_1$ and $\mathcal{B}_2$ do not commute:  $[\mathcal{B}_1, \mathcal{B}_2 ] = 2 ( |A_{down}\rangle \langle A_{up}| - |A_{up}\rangle \langle A_{down}| )$. Therefore, $W$ cannot assign a state to the lab and its contents that represents a definite outcome obtained by $A$, and at the same time describe the lab with the superposition $|\Phi^{+}\rangle =\frac{1}{\sqrt2} \left(|A_{up}\rangle + |A_{down}\rangle \right)$. 

So, as far as Wigner is concerned the laboratory is represented by a superposed state in which $A$ is entangled with the measuring device, and this excludes attributing to $A$ (or to $A$'s memory, or to the measuring device, etc.) one particular  definite property corresponding to the result $A$ obtained.\footnote{We assume the completeness of the standard formalism of quantum mechanics here.} So, even if Wigner is aware of all details of the experimental setup, and knows that his friend will perform a measurement inside her sealed laboratory, and is fully aware  that she will find a definite result, he still cannot take the  presence of one definite result as the basis for his own description of the lab and its contents. If he did, quantum mechanics tells us that his predictions would be falsified by the outcomes of his own future experiments.\footnote{See \cite{dieksfano} for a discussion of recent generalizations of the WF setup, in which two labs at a distance, with each an internal and an external observer, are considered. As argued there, these more complicated cases introduce questions about locality and Bell inequalities but do not shed new light on the perspectivalism issue.}

Instead of Alice and Wigner we could have considered inanimate physical systems; indeed, the only thing that counts in the above reasoning is the form of the relevant quantum states, which does not depend on $A$ and $W$ being agents in the ordinary sense of the word.\footnote{In the experiment as described an interaction takes place that correlates $A$ with the other systems in the lab. For $W$ the lab is represented by a superposition, even before any interaction between $W$ and the lab has occurred (the total state of $W$ and the lab is a product state). So, physical interactions are not absolutely necessary for the definition of a perspective; the existence of a relative state suffices, and this condition is also fulfilled in product states.}

The moral of the story is that the WF thought experiment shows that we need the concept of  \emph{relative} or \emph{perspectival} properties. The measuring device in the lab indicates a unique definite outcome relative to (or: from the perspective of) $A$, but there is no such determinate result relative to $W$. $W$ can verify, with experiments available to him in his situation, that his state attribution and the properties he assigns accordingly, is right. Similarly, $A$ possesses decisive experimental evidence for the presence of a definite $z$-spin value, even though there is no such value for $W$. This verifiability by means of standard physical procedures makes both descriptions \emph{objective} within their own perspectives. 

Generalizing this insight, we conclude that dynamic properties of physical systems are defined with respect to other physical systems. Acceptance of this perspectivalist doctrine entails a significant break with traditional reasoning about physical  properties: it implies that dynamic properties have to be represented by relational rather than monadic predicates. This  introduces a lack of ``universality'' of property attributions, since a property instantiated with respect to one system need not exist with respect to other systems. 

As already ponted out, this relativization of properties does not conflict with objectivity, in an important sense of that term. Both within $A$'s and $W$'s perspective there exists a standard distinction between true and false physical statements. For $W$ Alice's laboratory objectively possesses the properties associated with the superposition, and $W$ can justify his claim by measuring $\mathcal{B}_1$. It remains the case that what is thus objectively true for $W$ need not be true for $A$, and \emph{vice versa}. 

\subsection{A dilemma regarding the universality of facts}\label{dilemma}

A moment ago we suggested that there is in fact a stronger sense in which objectivity is ensured in quantum perspectivalism. If $W$ decides to learn what $A$ has found in her experiment, by measuring the observable $\mathcal{B}_2$, we submitted that the result of his measurement will faithfully represent what for $A$, in her perspective, was already the case. Intuition certainly tells us so: Wigner could just open the lab's door, enter and ask Alice what she has recorded. It seems obvious that an honest answer must reliably inform Wigner about $A$'s preexisting fact. If this is right, $A$'s and $W$'s perspectives make contact in this situation and come to share facts. These shared facts are objective in a stronger sense of objectivity than the one discussed above, namely ``acknowledged by different observers''. 

But on reflection it is not clear how this sharing of outcomes should follow from the quantum formalism. To see the problem, suppose that Alice has found ``spin up''. Still, Wigner predicts, on the basis of the state (\ref{eq:wignerstate}) that he has assigned to Alice and her laboratory, that with probability $1/2$ he will receive the answer ``spin down'' when he asks Alice about her result. Nothing in his calculation gives a special role to the result actually ascertained by Alice within her perspective, so nothing singles out Alice's result as the one Wigner must necessarily obtain. In repetitions of the experiment Wigner will therefore sometimes receive an answer that deviates from what Alice has registered. 

In other words, as far as standard quantum mechanics goes, Alice and Wigner live within their own independent perspectival worldviews \cite{brown,pienaar}, which need not have any overlap. If facts are nevertheless the same in their perspectives (\emph{e.g.}, if both Alice and Wigner find ``spin up'' for the outcome of the internal measurement), this is a matter of coincidence rather than a necessary consequence of a supposed universality of facts.

This isolation of different perspectives from each other does not lead to logical inconsistencies or conflicts with experience. Inspection of the state (\ref{eq:wignerstate}) shows that if Wigner hears from Alice that her result was ``down'', he will find confirmation of this answer in all records and traces in Alice's laboratory (everything from his own perspective, evidently). Even so, for Alice the outcome could be ``up'', which is also part of a consistent and verifiable system of empirical data, defined from her perspective. So, although it might seem that an ``ontology of relative facts is incompatible with scientific objectivity and undercuts the evidential base of quantum theory'' \cite{healeyrelativism}, this form of perspectivalism does not touch the empirical evidence for quantum mechanics and is not in conflict with a physically relevant notion of objectivity, open to empirical checks. It remains true, though, that such a lack of universality of facts, even if not detectable by any experiment, is highly counter-intuitive.      

The only way to meet our intuitive preferences here appears to \emph{stipulate} that there is, after all, a fact-sharing relation between different perspectives. This is what Healey does when he claims, in the context of his pragmatist interpretation \cite{healeyrelativism}, that Wigner, upon entering Alice's lab, will become part of Alice's local decoherence context and will therefore find the same result as Alice. Healey writes:
\begin{quote}
The applicability of a model of environmental decoherence licenses one to make the meaningful statement that this magnitude takes on some value, and to find out which. ... By relativizing truth to decoherence environments DP [\emph{i.e., Healey's proposed pragmatism}] is able to secure the immanent objectivity of scientific knowledge of the quantum world. This is because all actual quantum measurements may be assessed in a single shared context.
\end{quote}
But according to the quantum formalism both  ``spin up'' and ``spin down'' in equation (\ref{eq:wignerstate}) are part of a decoherence context, so decoherence by itself, as defined mathematically, does not suffice to privilege one of these outcomes as the one to be found by Wigner. It is only by \emph{assuming from the outset} that the local decoherence contexts occurring in measurements do not depend on perspectives but are shared among them that we may secure  ``immanent objectivity'', i.e.\ an objectivity requiring  universally valid facts. Obviously, the possibility of making this assumption of shared measurement results is not reserved to pragmatism: it can also be made within perspectival realism. 

We therefore end up with a case of interpretative underdetermination. There is the option of accepting the ``splendid isolation'' of each individual perspective, so that there are no common facts. The alternative is to stipulate that measurements produce local facts that are shared by different perspectives. In the latter case the perspectives of different observers locally measuring the same observable on the same physical system will agree about the measurement's outcome.

Application of these two interpretative options to the EPR situation will shed additional light on their merits.

\subsection{Perspectivalism and locality}\label{qpr3}

Suppose that Alice and Bob are space-like separated from each other and share a pair of spin-$1/2$ particles in the singlet state. This state can be represented by
\begin{equation}\label{singlet1}
\frac{1}{\sqrt{2}}(\ket{\up}_A \ket{\down}_B - \ket{\down}_A\ket{\up}_B),
\end{equation} 
where the labels $A$ and $B$ refer to Alice and Bob, respectively. Alice measures the spin of her particle with a device $M_A$, designed to measure spin in the $z$-direction.  

Alice obtains a definite result, ``up'', say. From Alice's perspective (or, from the perspective defined by $M_A$) her particle has therefore acquired the definite spin value ``up''.\footnote{In state (\ref{singlet1}) the coefficients are equal which implies rotational symmetry. This symmetry is at the basis of the notorious problem that no unique measurement basis  is singled out by states like (\ref{singlet1}). But in a less idealized description of the measurement, in which the many microscopic degrees of freedom of the macroscopic device plus those introduced by decoherence are taken into account this symmetry can be broken.} In addition, from $A$'s perspective Bob's particle now has a spin in the opposite direction. A perfect anti-correlation between the two particle spins was already present in the initial state (\ref{singlet1}), so what has changed is that a local event at $A$, namely ``up'' becoming definite, has locally affected $A$'s perspectival description of the world. It is true that this involves an aspect of quantum mechanics that is absent from classical mechanics: there is a well-defined correlation between $A$'s and $B$'s spins in state (\ref{singlet1}) even if there are no definite spin values in well-defined directions in this state. From all the possible instantiations of this correlation, one is selected, within $A$'s perspective, by $A$'s local measurement.

According to this account nothing changes at $B$'s position as a result of $A$'s measurement---this is analogous to the absence of an effect on Wigner of his friend's measurement.  By contrast, if it had been the case that the spin of Bob's particle in Bob's perspective had changed by Alice's measurement, this would certainly have signaled the presence of a non-local effect. 
But Bob, his particle and its spin do not feel an influence of what Alice is doing on her side of the experiment. Bob is sealed off from Alice and describes Alice and her measurement with unitary evolution; the result is a superposition in which both possible measurement results occur. The state used by Bob is  
\begin{equation}\label{singlet2}
\frac{1}{\sqrt{2}}(\ket{M_A \up}\ket{\up}_A \ket{\down}_B - \ket{M_A \down}\ket{\down}_A\ket{\up}_B),
\end{equation} 
with the two device states representing the outcomes ``up'' and ``down'', respectively. As required by the no-signaling theorem, the quantum state of Bob's particle (defined by its density matrix) has remained the same. If Bob is now going to perform a $z$-spin measurement on his particle, he predicts, on the basis of Eq.\ ({\ref{singlet2}) that there are two possible results, ``up'' and ``down''. So, if unitary quantum mechanics and the standard probability rules are empirically correct, Bob may find ``up'', even though from Alice's perspective her own particle had spin ``up'' and Bob's spin was ``down''.  

We might introduce a third observer, who describes both Alice and Bob unitarily, which results in the post-measurement state  
 \begin{equation}\label{singlet3}
\frac{1}{\sqrt{2}}(\ket{M_A \up}\ket{\up}_A \ket{\down}_B \ket{M_B \down}- \ket{M_A \down}\ket{\down}_A\ket{\up}_B\ket{M_B \up}).
\end{equation}  

According to the ``isolationist'' option from section \ref{qpr2}, the state (\ref{singlet3}) shows that the third observer will find a perfect anti-correlation between $A$'s and $B$'s results even though no such correlation need obtain between the results registered within $A$'s and $B$'s own perspectives. This illustrates the general situation: in the isolationist picture the outcomes of $A$'s and $B$'s measurements, as observed by a third experimenter, are independent from those outcomes in $A$'s and $B$'s perspectives. But the third observer will not be able to detect  these discrepancies. 

We may also ask what happens when $A$ and $B$ come together and compare notes. Both $A$ and $B$ predict (with certainty) that they will find a spin value registered at the other side that is opposite to the one found by themselves. Isolationism states that $A$ will find ``down'' as $B$'s registered outcome, and that $A$ will be able to verify that all traces left by $B$'s measurement agree with this outcome---even if from $B$'s viewpoint ``up'' was realized. A mirrored account applies to what $B$ will find out about $A$'s outcome. Within all perspectives the quantum predictions will be borne out so that all participants agree that quantum mechanics is supported by the empirical evidence.\footnote{A fuller discussion should also consider what happens when $A$ and $B$ measure in different directions, see \cite{dieksoxfhb}.}  

Evidently, this isolationist story lacks ``immanent objectivity'', defined as universality of measurement outcomes for all agents who observe the measurement. We may (with Healey) attempt to restore it by requiring that local measurement contexts are unique and define a single measurement result. In the case at hand this means that $A$, having traveled to $B$ and looking at his notebook, will see the result that $B$, from his perspective, wrote down when he made his measurement. If the quantum mechanical predictions are correct, this result must be opposite to what $A$ found in her own measurement. But this means that $A$'s measurement must have had an effect on $B$ and his surroundings, even though $B$ was located at space-like separation from $A$. If, in this account, $B$ also looks at $A$'s previous result, he will find the opposite of his own outcome, as expected.

The upshot is that there appears to be a trade-off between ``immanent objectivity'' and locality. The isolationist version of perspectivalism may be regarded as local, because nothing changes for $B$ when $A$ performs actions outside of his lightcone. According to this isolationism the world is a collection of completely independent perspectives. If a connection is forged between the different perspectives, in order to restore the universality of local measurement outcomes, a non-local global account results. 

\section{Locality in a fragmented world}

Let us for a moment return to QBism. As discussed in section \ref{qbism3}, QBists maintain the locality of their interpretation on the grounds that all quantum predictions pertain to local beliefs and expectations. As Fuchs, Mermin and Schack \cite{fuchsloc} write: 
\begin{quote}
when any agent uses quantum mechanics to calculate correlations between the manifold aspects of her experience, those experiences cannot be space-like separated. Quantum correlations, by their very nature, refer only to time-like separated events: the
acquisition of experiences by any single agent. Quantum mechanics, in the QBist interpretation,
cannot assign correlations, spooky or otherwise, to space-like separated events, since
they cannot be experienced by any single agent. Quantum mechanics is thus explicitly local
in the QBist interpretation.
And that’s all there is to it.
\end{quote}

Applying these ideas to a version of the EPR experiment in which Alice and Bob share a pair of particles in an entangled state, and in which both perform a measurement on their half of the total system, Fuchs, Mermin and Schack characterize the difference between their approach and the usual analysis as follows \cite{fuchsloc}:
\begin{quote}
What the usual story overlooks is that the coming into existence of a particular measurement
outcome is valid only for the agent experiencing that outcome. At the moment of his own measurement Bob is playing the friend to Alice’s far-away Wigner, just as at the moment of her own measurement she is playing the friend to Bob’s Wigner. Although
each of them experiences an outcome to their own measurement, they can experience an outcome to the measurement undertaken by the other only when they receive the other’s
report. Each of them applies quantum mechanics in the only way in which it can be applied, to account for the correlations in two measurement outcomes registered in his or her own
individual experience. And as noted above, experiences of a single agent are necessarily time-like separated. The issue of nonlocality simply does not arise.
\end{quote}

``Bob’s system is not changed by Alice’s far away
intervention in any way whatsoever'' \cite[p.\ 752]{fuchsloc}; it is only for Alice herself that her belief changes about what she will encounter in a future report from Bob. 

It is inherent in this QBist analysis that Bob may find the same result (valid within his private experience) as Alice became aware of. Indeed, before any measurements were made Bob could find ``up'' or ``down''; and nothing changes for Bob because of Alice's measurement. This in turn entails that Bob may find ``up'' and note this in his report, while Alice is absolutely certain, on the basis of her quantum calculation, that the report she will receive will say ``down''. If we do not want to question the empirical success of quantum mechanics, Alice's prediction-with-certainty about what she will find in Bob's report should come true (at least in the great majority of cases). In other words, what Alice reads in Bob's report is independent of Bob's experience of what he wrote down. Our actors have ``manifolds of experience'' that do not share facts.

It is not clear whether proponents of QBism are conscious of the fact that their interpretation leads to this disconnectedness of all experiential worlds. In their analysis of the WF paradox, Fuchs, Mermin and Schack write \cite[p.\ 751]{fuchsloc}: 
\begin{quote}
The disagreement between Wigner’s account and his friend’s is paradoxical only if you take a measurement outcome to be an objective feature of the world, rather than the contents
of an agent’s experience. The paradox vanishes with the recognition that a measurement outcome is personal to the experiencing agent. There is an outcome in the friend’s experience; there is none yet in Wigner’s. Of course their accounts differ. If Wigner goes on to ask his friend about her experience, then \emph{the disagreement is resolved the moment he receives her report, i.e. when it enters his own experience}. [Italics added.]
\end{quote} 

The italicized passage suggests that Fuchs, Mermin and Schack assume that the experiential worlds of Wigner and his friend will coincide as far as the result of the friend's measurement is concerned. But assuming this overlap between different experiential worlds, when they focus on the same measurement,  has the consequence that in EPR situations the presence of nonlocal influences has to be accepted, as explained in section \ref{qpr3}. This would spoil the QBist analysis of the EPR experiment, according to which everything is local.    

Therefore, the local QBist world must be a collection of independent agent-experiences. The alternative is to assume that agents share facts when they meet and interrogate each other; in this case Alice's measurement must have an influence at a distance, on Bob's possible experiences, so that the account becomes nonlocal.

Quantum perspectivalism replaces this agent-centered story with an account in which perspectives are centered on physical systems. Instead of Alice and Bob we could think of automated measuring devices $M_A$ and $M_B$, and when $M_A$ has registered a result, ``up'', say, the spin at the other side becomes ``down'', from the perspective of $M_A$. As argued before, this can be interpreted as a local process since the perfect anti-correlation existed already before the measurement  and the only change is the creation of a definite spin value by $M_A$. From $M_B$'s perspective nothing changes, so in a measurement $M_B$ will find ``up'' with probability $1/2$. When the worldlines of $M_A$ and $M_B$ cross, and the devices measure each other's outcomes, isolationism says that the two perspectives remain independent. The alternative is that $M_B$ must find, within its own perspective, the value predicted from $M_A$'s perspective. But this option makes the account nonlocal.

The local story does not allow perspective-independent objectivity and uses an infinity of independent perspectives in order to describe the totality of facts in the world. Although there is no conflict with experience, and although there is still objectivity internal to each perspective, this interpretative option may look extravagant.       
But it should be noted that very similar suggestions have recently been made in analytic metaphysics, under the heading of \emph{fragmentalism}. 

Fragmentalism was introduced by Kit Fine \cite{fine}, who claimed that it sheds new light on notoriously controversial questions surrounding tense and time flow, and was further investigated and elaborated by Lipman \cite{lipman0,lipman1,lipman2}. The central fragmentalist idea is that the world is not a monolithic whole constructed from mutually compatible facts, but rather a collection of \emph{fragments}, with each fragment containing mutually compatible facts, while different fragments are incompatible. Facts from different fragments conflict and therefore cannot co-obtain: different fragments cannot be combined into one consistent whole. According to fragmentalism, all fragments are needed for a complete description of the world, and they all enjoy the same status of objectively capturing an aspect of reality. The totality of reality is formed by the entire collection of all fragments. 

These ideas have been used to develop a fragmentalist analysis of the special theory of relativity \cite{lipmanrelativity}. In this analysis different reference frames supply different but equally real and objective facts, for example about lenghts, durations and simultaneity.  These facts are not ``mere appearances'' but each reference frame provides us with one consistent fragment of all facts obtaining in the world.

However, one may well object that in special relativity there is an absolute reality behind all these fragments, namely four-dimensional space-time with its Minkowski metric. The facts obtaining in this Minkowski space-time are mutually compatible, and on their basis all frame-dependent facts can be derived. So there seems little reason to attribute a fundamental status to the separate fragments containing frame-dependent facts. 

As we have seen, in unitary quantum mechanics there \emph{are} reasons  to think that there exist different descriptions of the world, corresponding to different perspectives, that cannot be embedded in one overarching view. Indeed, the local variant of perspectival quantum realism that we have discussed posits the existence of such independent collections of facts, relativized to different ``systems of reference''. If fragmentalism is considered to be a viable metaphysical position, potentially able to dissolve certain  metaphysical conundrums, it should certainly be taken seriously in the quantum context as well.\footnote{There has been an earlier attempt at a fragmentalist interpretation of quantum mechanics: Simon \cite{simon} has proposed that a superposition of quantum states corresponds to a collection of fragments of the physical world. This proposal faces serious difficulties, however \cite{iaqcalosi}.} 
This holds for local quantum perspectivalism, i.e.\ isolationism with its completely independent fragments, and it holds \emph{a fortiori} for the intuitively more palatable perspectivalism in which different perspectives share facts (as when Wigner enters the lab and finds the same result as his friend). 

However, as we have seen, the intuitive appeal of shared, perspective-independent measurement facts is offset by nonlocality in situations of the EPR type. In the context of unitary quantum mechanics nonlocal correlations lead to serious problems with relativistic covariance, as shown by various no-go theorems (see for a discussion of these problems and the drastic, counter-intuitive measures apparently needed to avoid them  \cite{dieksFoP2019}). So, from a theoretical point of view the radically fragmented world of local perspectivalism has its attractive sides after all.

\section{Conclusion}

The idea that the quantum world should be seen as consisting of a collection of perspectives, associated with possibly conflicting descriptions of physical systems and their properties, has gained a certain popularity because of its central importance to QBism, where these perspectives represent the subjective beliefs and expectations of agents. However, the arguments supporting perspectivalism do not hinge on the introduction of subjectivism in the interpretation of quantum mechanics. Perspectives can also be defined with respect to physical systems instead of agents. In fact, all single-world probabilistic interpretations of unitary quantum mechanics that pick out definite-valued quantities by an appeal to decoherence or a similar process need this kind of  perspectivalism to make sense of situations of the Wigner's friend type. These interpretations have the same resources as QBism, or Quantum Pragmatism, for handling conceptual puzzles like Wigner's friend and nonlocality; but they handle such puzzles in a perspectival realist way, without recourse to subjectivism. 
 
As it turns out, there are two such versions of perspectivalism. According to the first, when Wigner enters his friend's laboratory and asks her about her measurement result, he receives an answer that is independent of what his friend, from her own perspective, is aware of and has recorded in her notebook. In the second version it is stipulated that Wigner and his friend must  share the same outcome when they participate in the same local measurement context. 

The latter option may seem more plausible, but it leads to nonlocality and problems with relativistic covariance. The other version of perspectival quantum realism is local and can be made relativistically covariant, but it gives rise to a radical fragmentalism, according to which different perspectives are completely independent of each other and generally offer very different, and conflicting, descriptions of the world---though they are all needed for a total description of reality.

Remarkably, no physical experiment can enforce a decision  between these two interpretative options.

\section*{Declarations}
No funding was received for conducting this study and the author has no relevant financial or non-financial interests to disclose.

\end{document}